\documentclass[aps,floats,epsf, twocolumn]{revtex4}
 \usepackage{graphics}

\begin{document}

\title{Antilinear spectral symmetry and vortex zero-modes in topological insulators and graphene}
\author{Igor F. Herbut and Chi-Ken Lu}

\affiliation{Department of Physics, Simon Fraser University,
 Burnaby, British Columbia, Canada V5A 1S6}

\begin{abstract}

We construct the general extension of the four-dimensional Jackiw-Rossi-Dirac Hamiltonian that preserves the antilinear reflection symmetry between the positive and negative energy eigenstates. Among other systems, the resulting Hamiltonian describes the s-wave superconducting vortex at the surface of the topological insulator, at a finite chemical potential, and in the presence of both Zeeman and orbital couplings to the external magnetic field. Here we find that the bound zero-mode exists only when the Zeeman term is below a critical value. Other physical realizations pertaining to graphene are considered, and some novel zero-energy wave functions are analytically computed.

\end{abstract}
\maketitle

\vspace{10pt}

\section{Introduction}

Eigenstates with precisely zero energy which are localized in the core of topological defects in some order parameter in low-dimensional electronic systems have continued to draw attention. Historically, they were first studied in the context of fractionalization of electric charge \cite{jackiw, su, jr}. Nowadays, they are also thought to provide a possible route to fault tolerant quantum computation \cite{read,ivanov, nayak, wilczek}, as well as a  mechanism for ordering of the vortex core, for example \cite{herbut1, ghaemi, herbut2}. Their existence is often tied to index theorems, which in turn rely on the symmetry between positive and negative eigenvalues of the Hamiltonian. Such a reflection symmetry of the energy spectrum is certainly a necessary condition for the zero-energy states to be robust under variations in the defect potential: obviously, at least when their number is odd, at least one state has to be pinned to zero in order for the spectrum to remain symmetric under the exchange of the sign of the energy. This raises the following question: what is the general extension of the Jackiw-Rossi-Dirac \cite{jr} (JRD) Hamiltonian (defined below in Eq. (1)) which still possesses the symmetry in question? In this paper we provide the answer and discuss some physical realizations of our result.

We show that irrespectively of a representation, there are exactly two, one linear and one antilinear, operators which anticommute with the four-dimensional JRD Hamiltonian. The additional terms that respect the spectrum's reflection symmetry and which can be added to it then have to be odd under one of these two operators. An important example of a Hamiltonian that anticommutes with the linear operator is the JRD Hamiltonian in presence of the Abelian gauge field \cite{herbut4}. Here we focus on those terms that, in contrast, anticommute with the {\it antilinear} operator. In the four-dimensional representation there are precisely {\it four} such terms. Their physical meaning is maybe most transparent in the context of the superconducting vortex at the surface of the topological insulator \cite{fu}: 1) the chemical potential, 2) the Zeeman coupling of the electron spin to the external magnetic field, and 3) the two components of the electromagnetic vector potential. Some special cases of our general Hamiltonian have already appeared in literature in different physical contexts \cite{fu, babak, pi}. It is shown here that the most general Hamiltonian with the vortex of unit vorticity indeed has a zero-mode under a certain condition, and we exhibit the analytical solution for it when the flux of the gauge-field is localized at the origin. In the context of topological insulators, for example, we find that the Majorana zero-mode \cite{fu} exists only for the Zeeman term below a critical value, at which it delocalizes.

  The paper is organized as follows. In the next section we derive the antilinear operator that anticomutes with the JRD Hamiltonian. In Sec. III we display the extended JRD Hamiltonian consistent with the antilinear spectral symmetry, and discuss its physical realizations. In Sec. IV we solve analytically for the zero-mode of the spectrum. The important example of a vortex at the surface of the topological insulator in the magnetic field is discussed in Sec. V, and we summarize our findings in Sec. VI. Some simple, but for our purposes crucial algebraic facts about Dirac matrices are presented in the Appendix.

\section{Antilinear spectral symmetry}

The JRD Hamiltonian \cite{jr}  has the general form
\begin{equation}
H_0 = \alpha_1 p_1 + \alpha_2 p_2 + m_1 (\vec{x}) \alpha_3 + m_2 (\vec{x}) \alpha_4,
\end{equation}
where $\alpha_i$ are four Hermitian four-dimensional matrices that satisfy the Clifford algebra
$\{ \alpha_i, \alpha_j \} = 2 \delta_{ij}$, and the two masses are
$m_1 (\vec{x}) = 2 m(r) \cos n \phi$ and $m_2 (\vec{x})  = 2 m(r) \sin n \phi$, with $(r,\phi)$ as polar coordinates in the plane, and the integer $n$ is  the vorticity. We assume $m(r\rightarrow \infty)$ to be finite, but the function $m(r)$ otherwise arbitrary. $p_i$ are two components of the momentum operator. We have also set the velocity and the Planck constant to unity for convenience.

It is well-known that there is a unique constant matrix that anticommutes with all four matrices $\alpha_i$, and consequently with $H_0$: $\beta= \alpha_1 \alpha_2\alpha_3\alpha_4$, which we chose here to be Hermitian as well. The existence of this matrix implies that the spectrum is symmetric around zero, and $\beta$ plays a crucial role in the derivation of the index theorem for the JRD Hamiltonian \cite{weinberg}. There exists, however, a unique additional {\it antilinear} operator that anticommutes with $H_0$. To see this, the reader may recall that $H_0$ may be understood as the effective low-energy Hamiltonian for a non-uniform, but real tight-binding Hamiltonian on the graphene's honeycomb lattice \cite{hou}. Irrespectively of the representation one can thus always define an operation formally (and in the case of graphene, literally \cite{herbut3}) analogous to time-reversal symmetry: an {\it antilinear} operator that {\it commutes} with $H_0$. Such an effective time-reversal operator may then be written as $T= U K$, where $U$ is unitary, and $K$ stands for complex conjugation. The explicit form of $U$, of course, depends on the representation of the Clifford algebra. Since all four-dimensional representations are equivalent \cite{schweber, good}, on the other hand, one is at liberty to chose the most convenient one for the present purposes. Let us assume therefore a representation in which $\alpha_1$ and $\alpha_2$ are real, and $\alpha_3$ and $\alpha_4$ are imaginary. The existence of such a representation is proved in the Appendix. In this representation the unitary part of $T$ has to anticommute with all $\alpha_i$, and therefore there is a unique solution, $U=\beta$. The antilinear operator that anticommutes with $H_0$ in this representation is then simply
\begin{equation}
A= \beta T = K,
\end{equation}
where the first equality is completely general, and the second is specific to the representation. We have used the fact that $\beta^2=1$, and that the matrix $\beta$ in this representation must be real.

Uniqueness of the operator $A$ can also be seen to follow from Schur's lemma \cite{good}: since the unitary part of $A$ would have to commute with all four $\alpha$-matrices in the chosen representation, the irreducibility of the four-dimensional representation  guarantees that it is proportional to unit operator.

\section{General Hamiltonian}

It is easy then to identify all four-dimensional purely imaginary Hermitian matrices that as such anticommute with the operator $A$. The first class consists of those which are even under $T$ and anticommute with $\beta$: $\alpha_3$ and $\alpha_4$, already present in $H_0$, and $i\alpha_1 \beta= i \alpha_2 \alpha_3 \alpha_4$ and $i\alpha_2 \beta= -i \alpha_1 \alpha_3 \alpha_4$, which are not. The second class contains terms that are odd under $T$ and which commute with $\beta$: $i\alpha_1 \alpha_2$ and $i \alpha_3 \alpha_4$. The most general Hamiltonian, which while not anticommuting with $\beta$ any longer still anticommutes with the antilinear operator $A$, and as such has the reflection symmetry of the spectrum around zero, may thus be written as
\begin{widetext}
\begin{equation}
H= \sum_{i=1}^2 \alpha_i (p_i + A_i(\vec{x}) (i\alpha_3 \alpha_4)) + m_1 (\vec{x}) \alpha_3
+ m_2 (\vec{x}) \alpha_4  +  \mu(\vec{x}) i\alpha_3 \alpha_4 + h (\vec{x}) i\alpha_1 \alpha_2.
\end{equation}
\end{widetext}
It may also be useful to display the result using the standard covariant notation in terms of the Dirac (Hermitian) $\gamma$-matrices: $\alpha_i = i \gamma_0 \gamma_i$, for $i=1,2,3$, $\alpha_4 = i\gamma_0 \gamma_5$:
\begin{widetext}
\begin{equation}
H=  \sum_{i=1} ^2 i \gamma_0 \gamma_i (p_i + A_i(\vec{x}) \gamma_{35} ) + i \gamma_0 ( m_1 (\vec{x}) \gamma_3
+ m_2 (\vec{x}) \gamma_5)  +  (\mu(\vec{x})  + h (\vec{x}) \gamma_0) \gamma_{35}.
\end{equation}
\end{widetext}
The matrix $\gamma_{35}=i\gamma_3 \gamma_5$ that plays the prominent role in $H$ is the generator of the rotations of the two masses present in $H_0$ into each other. Since $\gamma_{35}$ has the eigenvalues $\pm 1$, $A_i (\vec{x})$ enters the Hamiltonian as the axial gauge field, although we will see that it may also represent the true electromagnetic field in some physical realizations, notably in the topological insulator. $h(\vec{x})$ is the mass-term which preserves the chiral symmetry of the massless Dirac Hamiltonian, generated by $\{\gamma_3, \gamma_5, \gamma_{35} \}$, but breaks the (effective) time-reversal symmetry.

 The Hamiltonian in Eq. (4) represents the most general extension of the JRD Hamiltonian that preserves the antilinear symmetry between the positive and negative parts of the energy spectrum, with the additional terms independent of  momentum. Momentum-dependent extensions are also physically relevant, and will be a subject of a separate publication.\cite{ckl} In this paper we have not considered the extensions of the JRD Hamiltonian that would preserve the linear spectral symmetry provided by $\beta$ either. As already mentioned, an important example of the latter is the addition of the Abelian gauge potential by minimal subtraction, $p_i \rightarrow p_i - A_i$, which represents graphene in magnetic field, for example, and also yields zero-energy states \cite{herbut4}.

 Special cases of the Hamiltonian $H$ have already arisen in different physical contexts. When $\mu(\vec{x})= h(\vec{x})=0$, Jackiw and Pi \cite{pi} have found that the axial vector potential can in certain sense be factored out of the Hamiltonian, so that its presence does not alter the number of zero-energy states, but only modifies their form. When $A_i (\vec{x})=h (\vec{x})=0$, the Hamiltonian describes the vortex in the s-wave superconducting order parameter at the surface of the topological insulator at a finite chemical potential \cite{fu, bergman, khyamovich}, and the exciton condensate in the symmetrically biased graphene bilayer \cite{babak}. It may also be understood as describing the vortex in the N\' eel order parameter at finite Zeeman coupling to magnetic field in graphene, where one needs one copy of $H$ for each Dirac point \cite{herbut1}. Similarly, the Bogoliubov - de Gennes Hamiltonian for the vortex in a general superconducting order parameter on graphene's honeycomb lattice with the orbital effect of the magnetic field included requires two copies of $H$ with $h (\vec{x})=0$ \cite{herbut2}.

\section{Zero-mode }

 Let us now proceed to solve for the zero-energy state for $\mu(\vec{x})=2\mu $ and $h(\vec{x})=2h$ constant, when an analytic solution is possible. In graphene, for example, we may chose $\alpha_1= -\sigma_3 \otimes \sigma_1$, $\alpha_2 = I\otimes \sigma_2$, $\alpha_3=  \sigma_1 \otimes \sigma_1$, and $\alpha_4 = \sigma_2 \otimes \sigma_1$ \cite{herb-jur-roy}. The Dirac fermion is then $\Psi^\top = (u_1, v_1, u_2, v_2)$, with $u_i$ and $v_i$ standing for the components of the wave-functions at the two triangular sublattices of the honeycomb lattice, near the two Dirac points. (True spin degree of freedom is suppressed.) The masses are chosen so to represent the vortex in the Kekul\' e bond-density-wave \cite{hou}. The term proportional to $\mu$ is in this example the ``pseudo-chemical potential", which has the opposite sign at the two Dirac points, whereas the one proportional to $h$ represents the pattern  of imaginary-valued hoppings between the sites belonging to the same sublattice \cite{haldane}. The vector potential may be understood as arising from a ripple on the graphene's surface, for example \cite{herbut3}. The equations for the zero-energy state in this representation become
 \begin{equation}
 (\mu + h) u_1 + (i \partial_{\bar{z}} -A) v_1 + \bar{m} v_2 =0,
 \end{equation}
 \begin{equation}
 -(\mu + h ) u_2 + m v_1 -(i \partial _z +\bar{A})   v_2 =0,
 \end{equation}
 \begin{equation}
 (\mu -h ) v_1 + (i \partial _z  -\bar{A})    u_1 + \bar{m}   u_2=0,
 \end{equation}
 \begin{equation}
 -(\mu -h ) v_2  +m  u_1  -  (i \partial_{\bar{z}}+A)    u_2 =0,
 \end{equation}
 where $m= m(r) e^{i\phi}$, $\partial _z= e^{-i\phi} (\partial_r - (i/r) \partial_\phi)$, $A= A_1 + i A_2$, and the bar denotes a complex conjugation. Assuming an {\it ansatz}
\begin{equation}
e^{  i (\frac{\pi}{4} - \phi)  (-)^k } v_k = (-)^{k+1} g(r) e^{-\int_0 ^ r m(r') dr'},
\end{equation}
\begin{equation}
e^{  i \frac{\pi}{4}  (-)^{k+1} } u_k =  f(r) e^{-\int_0 ^ r m(r') dr'},
\end{equation}
with $k=1,2$, Eqs. (5)-(8) reduce to only two:
\begin{equation}
(\mu+h) g (r)+ ( \partial_r + A_\phi (r))    f(r) =0,
\end{equation}
\begin{equation}
(\mu-h) f(r) - (\partial_r + \frac{1}{r} - A_\phi (r) ) g(r) =0,
\end{equation}
where we have also assumed a spherically symmetric field gauge field $ \vec{A}(\vec{x}) = A_\phi (r) \hat{\phi}$.

\subsection{Zero gauge-field}

Let us first consider the problem without the gauge field, $A_\phi(r) \equiv 0$. Combining the last two equations, one finds the standard Bessel differential equation for the function $g(r)$:
\begin{equation}
r^2 \partial^2 _r g(r) + r \partial_r g(r) + [(\mu^2 - h^2)r^2  - 1] g(r) =0.
\end{equation}
For $|h|< |\mu|$ the solution normalizable at the origin is qualitatively similar to the solution for $h=0$ \cite{bergman, khyamovich}:
\begin{equation}
g(r)= C J_1 ( r (\mu^2 -h^2)^{1/2} ),
\end{equation}
\begin{equation}
f(r)= C   ( \frac{ \mu +h}{\mu -h} )^{1/2}      J_0 (r (\mu^2 -h^2)^{1/2} ),
\end{equation}
where $J_n (z)$ are the Bessel functions of the first kind, and $C$ is the normalization constant. The only difference from the solution at $h=0$ is in the characteristic length scale of oscillations, which now became longer. As long as $|\mu| \geq |h|$ the zero-energy state is exponentially localized far from the vortex, with the characteristic length scale $\sim 1/m(\infty)$. At $\mu=h$ two of the components vanish identically, $v_1=v_2 =0$, while the other two become constant. The zero-mode in this limit becomes the same as the one of the original JRD Hamiltonian \cite{jr}. At the opposite end when $\mu=-h$, however, the solution is different, $f(r)=C $, $g(r)= C \mu r$. In either case, at $|\mu|=|h|$ the characteristic length scale for the oscillations under the overall exponential decay of the solution diverges.

When $|\mu|<|h|$, on the other hand, the solutions turn into the  modified Bessel functions
\begin{equation}
g(r)= C I_1 ( r (h^2-\mu^2)^{1/2} ),
\end{equation}
\begin{equation}
f(r)= - C ( \frac{ \mu +h }{ h- \mu } )^{1/2} I_0 ( r (h^2-\mu^2)^{1/2}  ),
\end{equation}
which now at large radius {\it grow} exponentially,
\begin{equation}
g(r) \propto \frac{e^{r (h^2-\mu^2)^{1/2} }}{r (h^2-\mu^2)^{1/2}  }.
\end{equation}
The zero-energy state therefore remains normalizable only if the following condition is met:
\begin{equation}
m(\infty)^2 + \mu^2 \geq h^2.
\end{equation}
When $|h|$ is above the critical value the zero-mode is exponentially large far from the vortex. In a finite system this would presumably correspond to the state becoming localized at the boundary. Right at the critical value the zero-mode is thus {\it critical}: it oscillates with the amplitude of oscillations decaying as a power-law $\sim 1/\sqrt{r}$.

\subsection{Finite gauge-field}

  With the general gauge-field present it seems not to be possible any more to find the analytic solution for the zero-mode, in contrast to the case when $\mu=h=0$ \cite{pi}. We can nevertheless solve analytically a somewhat artificial but nevertheless instructive example  of the field's strength as the delta-function at the origin, when $A_\phi(r) = \Phi/2r$. To determine the sign and the magnitude of the flux $\Phi$ that would render the energy of the vortex configuration assumed here finite, note that the Hamiltonian $H$ in Eq. (4) is invariant under the local unitary transformation,
  \begin{equation}
  H\rightarrow e^{i \theta(\vec{x}) \gamma_{35}}  H  e^{- i \theta(\vec{x}) \gamma_{35}},
  \end{equation}
  provided it is accompanied by the gauge transformation
  \begin{equation}
  A_k \rightarrow A_k + \partial_k \theta(\vec{x}),
  \end{equation}
  and the rotation of the complex mass,
  \begin{equation}
  m \rightarrow m e^{2i \theta(\vec{x})}.
  \end{equation}
  The last two transformations imply that the gauge-invariant coupling of the mass and the vector potential has the form
  \begin{equation}
  |(\partial_k - 2i A_k)m|^2 ,
  \end{equation}
  so that the vortex configuration in the mass $m$ has the finite energy only if $A_\phi = 1/2r$ at large distances. We thus take the flux to be
  $\Phi = +1/2$. Equations (11) and (12) then become quite similar:
  \begin{equation}
(\mu+h) g (r)+ ( \partial_r + \frac{1}{2r} )    f(r) =0,
\end{equation}
\begin{equation}
(\mu-h) f(r) - (\partial_r + \frac{1}{2r} ) g(r) =0.
\end{equation}
For $|\mu|>|h|$ the solutions are still the Bessel functions of the first kind, but now of the order $\pm 1/2$:
\begin{widetext}
\begin{equation}
g(r) = C_1 J_{1/2} (r (\mu^2 - h^2)^{1/2} )  + C_2  J_{-1/2} (r (\mu^2 - h^2)^{1/2} ),
\end{equation}
\begin{equation}
f(r) = ( \frac{\mu+h}{\mu-h})^{1/2} [C_1 J_{1/2} (r (\mu^2 - h^2)^{1/2} )  - C_2  J_{-1/2} (r (\mu^2 - h^2)^{1/2} )].
\end{equation}
\end{widetext}
One of the two constants appearing in the solutions is to be fixed by the condition at the origin, which, as usual, must be additionally specified \cite{juricic, melikyan}, and which would correspond to different short-distance regularizations of the magnetic flux.

Spreading the flux over a finite region of a linear size $\sim \lambda$ around the origin may be accomplished by defining $A_\phi = (1/2 \lambda^2 ) r$ for $r<\lambda$, and $A_\phi = 1/2r$, for $r>\lambda$, for example. Very near the origin then the differential equation would reduce to Eq. (13) with $\mu^2 \rightarrow \mu^2 + (1/2 \lambda^2)$, which would still yield one regular solution. At $r=\lambda$ the continuity of the solution would place two constraints on three (one for $r<\lambda$, and two for $r>\lambda$) constants of integration. The zero-mode clearly still exists. When $|h|>|\mu|$,  modified Bessel functions replace those of the first kind far from the origin, and the zero-mode again delocalizes when the condition in Eq. (19) ceases to hold.

\section{Discussion}

The graphene representation we used is such that $\alpha_i$ are real for $i=1,3$ and imaginary for $i=2,4$. The  time-reversal operator is then $T=i\alpha_1 \alpha_4 K$, and therefore $A= i\alpha_2 \alpha_3 K = \sigma_1 \otimes \sigma_3 K $. The zero-energy mode we found is therefore an eigenstate of the operator $A$ with the eigenvalue $+1$. It is easy to check that the other eigenstate of $A$ with the eigenvalue $-1$ has the opposite sign of the exponential in Eqs. (9) and (10), and therefore is not normalizable.

The antilinear operator that anticommutes with the special case of our Hamiltonian when $h=0$ was already recognized in the second of the Ref. 14, in a specific representation similar to the one for graphene. The readers familiar with the BCS theory of superconductivity may also recognize it as being closely related to the  ubiquitous symmetry of the BCS-type Hamiltonians, which originates in the Bogoliubov-Valatin doubling of degrees of freedom, characteristic for the BCS problem. We saw here, however, that the antilinear spectral symmetry is in fact a general property of certain wide class of Dirac Hamiltonians, which derives from the properties of the representations of Clifford algebra under complex conjugation. As such it is also present in the realizations of the Dirac Hamiltonian relevant to graphene, where the masses may equally represent {\it insulating} order parameters \cite{herb-jur-roy, ryu}.

At the surface of the topological insulator, we can consider the BdG Hamiltonian in presence of the vortex in the s-wave superconducting order parameter \cite{fu, linder}, by constructing the Dirac fermion as
$\Psi ^\top = (c_{\uparrow}, c_\downarrow, c^\dagger _\uparrow, c^\dagger _\downarrow) $. If the single particle Hamiltonian is $h= p_1 \sigma_1 + p_2 \sigma_2 -\mu$, the BdG Hamiltonian at $\mu=0$ assumes the form in Eq. (1) but with the $\alpha$-matrices as $\tilde{\alpha} _1 = I\otimes \sigma_1$,  $\tilde{\alpha}_2 = \sigma_3 \otimes \sigma_2$,  $\tilde{\alpha}_3 = \sigma_1 \otimes \sigma_2$,  $\tilde{\alpha}_4 = \sigma_2 \otimes \sigma_2$. Although this representation appears rather different from the one we used, the two are, of course, equivalent: $\tilde{\alpha}_i = U^\dagger \alpha_{i} U$,  for $i=1,2,3,4$, with the unitary operator as
\begin{equation}
U= e^{i \frac{\pi}{4} (\sigma_3 \otimes \sigma_3) } [ I\otimes e^{i \frac{\pi}{4} \sigma_3 } ] [ \sigma_3 \otimes \sigma_2].
\end{equation}
Since the matrix $-\sigma_3\otimes I= i\tilde{\alpha}_3\tilde{\alpha}_4$ is the particle number operator, both the chemical potential and the electromagnetic field enter the BdG Hamiltonian by multiplying it, just like in Eq. (3). Finally, as the generator of spin rotations around the direction perpendicular to the plane of the system is $ \sigma_3 \otimes \sigma_3 = -i \tilde{\alpha}_1 \tilde{\alpha}_2$, the Zeeman coupling of the electrons to the magnetic field of the vortex, enters the BdG Hamiltonian precisely as the last term in Eq. (3).

The superconducting vortex at the surface of the topological insulator provides therefore a physical realization of the most general Hamiltonian with the antilinear reflection symmetry of its spectrum. As the magnetic field always accompanies the superconducting vortex, we conclude that the Majorana fermion in the vortex will survive only if the Zeeman coupling of the electrons in the topological insulator to  the magnetic field is small enough. In this respect we may note that our assumption of a constant Zeeman term is a reasonable approximation in the strong type-II limit, in which the magnetic field decays over the length scale of penetration depth $\lambda$, whereas the zero-energy state decays over the much shorter superconducting coherence length, $\xi \sim 1/m(\infty)$. According to Eq. (19), however, even for an overly strong Zeeman coupling the Majorana fermion can always be produced inside the vortex by simply increasing the chemical potential. Finally, one can imagine placing the whole system in an additional uniform magnetic field, which can then be used to manipulate the localization of the Majorana zero-mode.

\section{Conclusion}

In conclusion, we have determined the general extension of the four-dimensional Jackiw-Rossi-Dirac Hamiltonian that retains the antilinear reflection symmetry of the spectrum, and solved for its zero-energy state in several examples. A particularly relevant physical realization of the most general Hamiltonian of this kind is provided by the superconducting vortex at the surface of a topological insulator with the vortex, and/or an external magnetic field fully included. The Majorana fermion inside the vortex core is found to exist only when the Zeeman coupling of electrons to the magnetic field is sufficiently small.

\section{Acknowledgement}

This work has been supported by the NSERC of Canada (IFH) and the NSC of Taiwan (CKL). We thank B. Seradjeh for useful discussions.

\section{Appendix}

If one demands that five four-dimensional anticommuting Hermitian matrices, such as  $\alpha_i$ $i=1,..4$ and $\beta$, are all either real or imaginary, it is easy to see that in any representation precisely three of these will be real and two imaginary.

A direct way to show this is to  construct the matrices $\alpha_i $ and $\beta$ out of standard Pauli matrices. The first three one may chose to be
  \begin{equation}
  \sigma_k \otimes \sigma_1,  \sigma_k \otimes \sigma_2, \sigma_k \otimes \sigma_3.
  \end{equation}
  The remaining two will then be
  \begin{equation}
  \sigma_n \otimes I, \sigma_m \otimes I
  \end{equation}
  where $n\neq m\neq k$. Choosing $k=1$ or $k=3$ then makes the second in the first group and one in the second group imaginary, and the rest real. Choosing $k=2$, on the other hand, makes the first and the third in the first group imaginary, and the rest real. Since all sets of five four-dimensional anticommuting Hermitian matrices of definite symmetry under complex conjugation are either exactly like in this example, or with the the two factor spaces interchanged, this proves our assertion.

    Another way to prove it would be to notice first that it would be impossible to have three of the matrices imaginary and two real, since that would contradict the fact that any of the five matrices is a product of the remaining four. So {\it a priori} the only other options would be to have all five  matrices real, or four imaginary and one real. Both cases would imply that there exists a four-dimensional representation of the Clifford algebra of four elements that all square to +1 or all to -1,  which is purely real, since in the latter case we could pick the four imaginary matrices and multiply them by the imaginary unit. But that would, on the other hand, be in  contradiction with the result that the smallest real representations of Clifford algebras $C(4,0)$ (four anticommuting elements each squaring to +1) and $C(0,4)$ (four anticommuting elements each squaring to -1) is actually  eight-dimensional \cite{okubo}. The only Clifford algebras of four elements that actually possess a real four-dimensional representation are $C(3,1)$ and $C(2,2)$, in accord with our result.

  In our derivation of the antilinear symmetry of the JRD Hamiltonian we could indeed therefore pick two real matrices for those multiplying the momentum operator, and the two imaginary for the masses.

\end{document}